\documentclass[useAMS,usenatbib,usegraphicx]{mn2e}

\title[Cold Flows and Reservoirs]{The Role of Cold Flows and Reservoirs in Galaxy Formation With Strong Feedback}
\author[R.~M.~Woods et al.]{R.~M.~Woods$^{1}$\thanks{E-mail:
woodsrm@mcmaster.ca}, J.~Wadsley$^{1}$, H.~M.~P.~Couchman$^{1}$, G.~Stinson$^{2}$ and S.~Shen$^{3}$\\
$^{1}$Department of Physics and Astronomy, McMaster University, Hamilton,
Ontario L8S 4M1, Canada\\
$^{2}$Max-Planck-Institute f\"ur Astronomie, K\"onigstuhl 17, D-69117 Heidelberg, Germany\\
$^{3}$Department of Astronomy, University of California, Santa Cruz, CA 95064, USA}

\begin{document}
\bibliographystyle{mn2e}

\date{\today}

\pagerange{\pageref{firstpage}--\pageref{lastpage}} \pubyear{2013}

\maketitle

\label{firstpage}

\begin{abstract}
{We examine gas accretion and subsequent star formation in representative galaxies from the McMaster Unbiased Galaxy Simulations \citep{stinson10}. Accreted gas is bimodal with a natural temperature division at $10^5$ K, near the peak of the cooling curve. Cold-mode accretion dominates inflows at early times, creating a peak in total accretion at redshift z=2-4 and declining exponentially below z$\sim$2. Hot-mode accretion peaks near z=1-2 and declines gradually. Hot-mode exceeds cold-mode accretion at z$\sim$1.8 for all four galaxies rather than when the galaxy reaches a characteristic mass.  Cold-mode accretion can fuel immediate star formation, while hot-mode accretion preferentially builds a large, hot gas reservoir in the halo.  Late-time star formation relies on reservoir gas accreted 2-8 Gyr prior.  Thus, the reservoir allows the star formation rate to surpass the current overall gas accretion rate.  Stellar feedback cycles gas from the interstellar medium back into the hot reservoir.  Stronger feedback results in more gas cycling, gas removal in a galactic outflow and less star formation overall, enabling simulations to match the observed star formation history. For lower mass galaxies in particular, strong feedback can delay the star formation peak to z=1-2 from the accretion peak at z=2-4.}
\end{abstract}

\begin{keywords}
methods: numerical - galaxies: evolution - galaxies: formation - cosmology: theory.
\end{keywords}

\section{Introduction}
\label{sec:introduction}

The formation of luminous galaxies depends on the accretion of gas. How that gas accretion changes as the galaxy grows in mass will determine many important properties of the galaxy. The traditional view of galaxy formation stems from a number of early theoretical papers, \citet{reesOstriker77}, \citet{silk77} and \citet{fallEfstathiou80}.  In this model, spherically infalling gas shock heats to the virial temperature near the virial radius, creating a thermally supported halo of gas. In the inner regions of the halo, gas is able to radiate away much of its thermal energy, and cool and collapse down to a centrifugally supported disc where star formation may occur.

Semi-analytic models of galaxy formation typically begin with this picture \citep{whiteFrenk91, kauffmann93, cole94, avila-reese98, mo98, somervillePrimack99, stringerBenson07, kampakoglouSilk07}, in which a density profile is assumed and a cooling radius is then calculated, specifying the point at which gas is able to collapse.

There is a growing body of work examining the robustness of shocks near the virial radius. \citet{dekelBirnboim06} argue that in order for gas to shock, the compression rate of the infalling gas must be larger than the cooling rate.  Smaller haloes, with short cooling times, would rapidly radiate away most of the compression energy and thus only support an isothermal shock at small radii with no extended hot halo.

Most galaxies live at the nodes of the filamentary structure of dark matter (DM) in our Universe \citep{bond96}. Many authors have noted that this enables gas to effectively be funnelled into the galaxy along the DM filaments \citep{katz93, katzWhite93, katz94, bertschingerJain94, bond96, shen06, harford08}. It has only been in the past decade that simulations have shown that these filaments may also be a means for gas to get into a galaxy without shock heating \citep{birnboimDekel03, keres05, dekelBirnboim06, keres08, ocvirk08, dekel09, brooks09, stewart11a}, {but resolving the details of cold flows is numerically challenging and is the subject of ongoing work \citep{nelson13}}.  Gas that is able to accrete without shocking at the virial radius is generally referred to as  `cold-mode accretion', while gas that penetrates along the filaments is called a `cold flow' (a type of cold-mode accretion; \citealt{keres05}).

Recent papers have suggested that signatures of cold-mode accretion, and specifically of cold flows, should be visible in absorption-line studies of galaxies, especially for Mg \textsc{II} systems \citep{stewart11a, kimm11}, which are optically thick in neutral hydrogen \citep{rigby02}, the primary component of cold-mode gas. As well, it has been suggested that cold flow gas has very high specific angular momentum,  and should be visible as a radial velocity offset, distinct from outflowing gas \citep{stewart11b}. However, observational evidence for cold flows in galaxies is currently limited (e.g. \citealt{steidel10}). For all but the closest galaxies, resolving cold flow features is very difficult due to a low covering fraction of the flows on the galaxy \citep{faucherKeres11}.  Though observations of cold-mode accretion have already been made \citep{stewart11b, crighton13}, the widespread existence of cold flows is yet to be established observationally.

{The mode of gas accretion is very important to our understanding of galaxy formation because it determines when accreted gas will be available to form into stars. Shock-heated gas generally ends up in an area of phase space with long cooling times, and so the gas may not become available for star formation until long after the gas was actually incorporated into the halo. Conversely, gas which is able to stay cold during accretion can fuel star formation almost immediately. Thus, instantaneous star formation rates (SFRs) depend not only on the current gas accretion rates, but also on previous accumulation of gas and on the mode of that accreted gas.} For this reason, further characterizing and understanding of accretion modes is important for getting a better picture of galaxy formation.

In this paper, we seek to characterize the gas accretion and star formation histories in simulations of four different representative galaxies with a range of masses near $10^{12}$ $M_{\odot}$.  The goal is to determine how accretion mode and feedback affect star formation in typical galaxies with masses similar to the Milky Way.  This work is one of a number of studies using some of the same galaxies.  For example, \cite{brook13} examined the metallicity, but not the mode of the accreting gas, with a strong feedback model.  By exploring both weak and strong feedback models, we are able to separate out behaviours that are more generic, such as total accretion, and those that are more dependent on feedback, such as the detailed star formation history.  In section \ref{sec:method}, we describe our sample of simulated galaxies as well as our tracking methods and gas-mode classification scheme. Section \ref{sec:results} details how the different accretion modes contribute to total gas accretion and feed star 
formation and how feedback changes this.

\section{Methods}
\label{sec:method}

\subsection{MUGS}

We examine how gas is accreted on to a selection of galaxies from the McMaster Unbiased Galaxy Simulations (MUGS).  MUGS is a suite of simulations designed for understanding aspects of galaxy formation through examination of an unbiased sample of galaxies simulated at high resolution \citep{stinson10}. To date, 16 different galaxies have been simulated within a mass range of $4.5 \times 10^{11}$ - $2\times10^{12} M_{\odot}$ using the Tree-SPH code \textsc{gasoline} \citep{wadsley04}. Physics in the MUGS runs include metal cooling \citep{shen10b}, UV background heating, star formation and stellar feedback using the approach of \citet{stinson06}.

MUGS galaxies are high-resolution re-simulations of 1 Mpc regions drawn from a 50 h$^{-1}$ Mpc$^{3}$ cosmological volume based on Wilkinson Microwave Anisotropy Probe 3 (\textit{WMAP}3) cosmology. Cosmological parameters were taken from the \textit{WMAP}3 $\Lambda$ cold dark matter cosmology \citep{spergel07}. Table \ref{tab:galaxysample} shows that the virialized region of each galaxy is resolved with $\sim10^6$ particles. More detailed information can be found in \citet{stinson10}.

For the purposes of this study, four galaxies were selected from the MUGS runs, details of which are presented in table \ref{tab:galaxysample}. The selected galaxies span the mass range available within the MUGS sample and have many observational properties consistent with observed galaxies \citep{stinson10}.  This agreement is a product of the high-resolution and detailed physical models employed.  However, this does substantially increase the simulation cost per galaxy and thus limits the sample size. Three of the four galaxies (g7124, g1536 and g15784) were also re-simulated using the stronger feedback scheme present in the MAking Galaxies In a Cosmological Context (MAGICC) study \citep{stinson13}. These galaxies are listed with the `magicc' suffix in table \ref{tab:galaxysample} and with an `m' suffix in plots. The stronger feedback includes substantial pre-supernova, early stellar feedback and has a higher fraction of massive stars so that the total energy per unit stellar mass is much higher.  We refer the reader to \citet{stinson13} for details.  The increased feedback in the MAGICC model produces galaxies that better match the baryon content and star formation histories of observed galaxy samples, such as the results of \citet{moster13}.  This leads to better rotation curves, surface brightness profiles and H \textsc{I} gas distributions \citep{stinson13}.  The stronger feedback model is definitely at the high end in terms of absolute energy input.  Inevitable losses due to finite resolution, such as those associated with mixing and cooling that could be avoided at higher resolution, lower the effective feedback.  The final result is remarkably consistent with many observations.  In this study, we look upon the strong (MAGICC) feedback and the original MUGS feedback as upper and lower bounds, respectively, effectively bracketing the true effect of stellar feedback.  In addition, comparing against the stronger feedback model allows inferences to be drawn on the role of feedback in creating delays between accretion of gas and formation of stars.   This study thus provides direct insight into how stronger feedback is able to reshape the star formation history to match observational expectations.

\begin{table}
\begin{center}
\begin{tabular}{ccccc}
Galaxy & Mass (M$_{\odot}$)  & $N_{gas}$ & $N_{star}$ & $N_{dark}$\\ \hline \hline
g1536 & 7.0$\times 10^{11}$& 2.4$\times 10^5$ & 1.4$\times 10^6$ & 5.3$\times 10^5$\\
g1536-magicc &  6.6$\times 10^{11}$ & 3.0$\times 10^5$ & 6.0$\times 10^5$ & 5.3$\times 10^5$\\
g7124 & 5.0$\times 10^{11}$ & 1.4$\times 10^5$ & 1.2$\times 10^6$ & 3.7$\times 10^5$\\
g7124-magicc & 4.5$\times 10^{11}$ & 1.9$\times 10^5$ & 1.6$\times 10^5$ & 3.6$\times 10^5$\\
g15784 & 1.5$\times 10^{12}$ & 5.3$\times 10^5$ & 2.6$\times 10^6$ & 1.2$\times 10^6$\\
g15784-magicc & 1.2$\times 10^{12}$ & 4.7$\times 10^5$ & 2.2$\times 10^6$ & 9.3$\times 10^5$\\
g15807 & 2.3$\times 10^{12}$ & 8.7$\times 10^5$ & 4.0$\times 10^6$ & 1.7$\times 10^6$\\
\hline
\end{tabular}
\end{center}
\caption[Sample of z=0 galaxies used from MUGS for analysis.]{Sample of z=0 galaxies used from MUGS for analysis. $N_{x}$ is the number of particles of type $x$ contained within the virial radius.}
\label{tab:galaxysample}
\end{table}

\subsection{Analysis}

In order to characterize the history of accreted gas in galaxies, gas particles are tracked throughout their history in the simulation. Haloes are identified at each output using the Amiga Halo Finder (\textsc{AHF}; \citealt{knollmanKnebe09}). We begin tracking by identifying all gas particles that formed stars and that were ever part of the main halo throughout the history of the simulation. We also identify all gas that is ever part of the main halo, regardless of whether or not it forms stars. All of these gas particles are then tracked backwards in time all the way back to the earliest time that \textsc{AHF} is able to identify haloes (roughly redshift 11). Tracking both the specific gas particles and the haloes enables us to determine the mode of accretion even at early times when the main halo may have consisted of many progenitors. Note that for stars, tracking for the parent gas particle starts at the formation time of that star. Gas that does not form a star is tracked at every snapshot for which it is 
available.

\subsection{Accretion categories}

\begin{figure}
 \includegraphics[natwidth=800,natheight=825,width=0.5\textwidth]{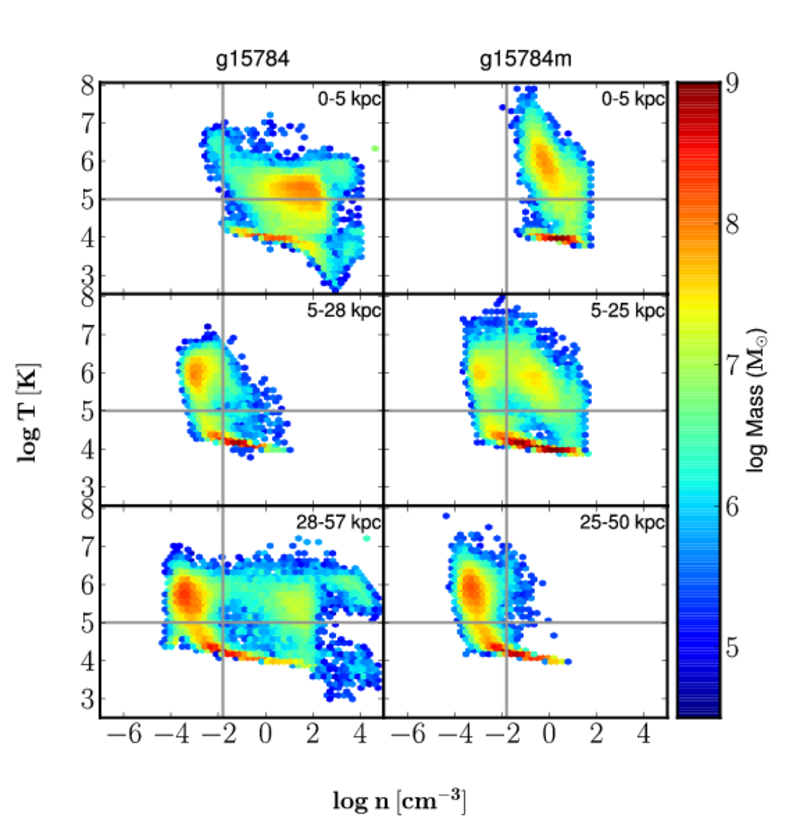}
  \caption{{Phase diagrams at z = 3 for g15784 (left column) and g15784m [right column, the high-feedback scheme of \citet{stinson13}]. The three rows from top to bottom represent radial shells from 0-0.1, 0.1-0.5 and 0.5-1 $r_{\mbox{vir}}$, respectively. Note that there is relatively little material near the two thresholds of $10^5$ K and $0.01$ cm$^{-3}$, drawn as grey lines. Data have been binned into regular hexagons of wall-to-wall width 0.24 in log $\rho$ space and height $\sqrt{3}$ width in log $T$ space.}}
  \label{fig:phaseshell}
\end{figure}

\begin{figure}
 \includegraphics[natwidth=800,natheight=550,width=0.5\textwidth]{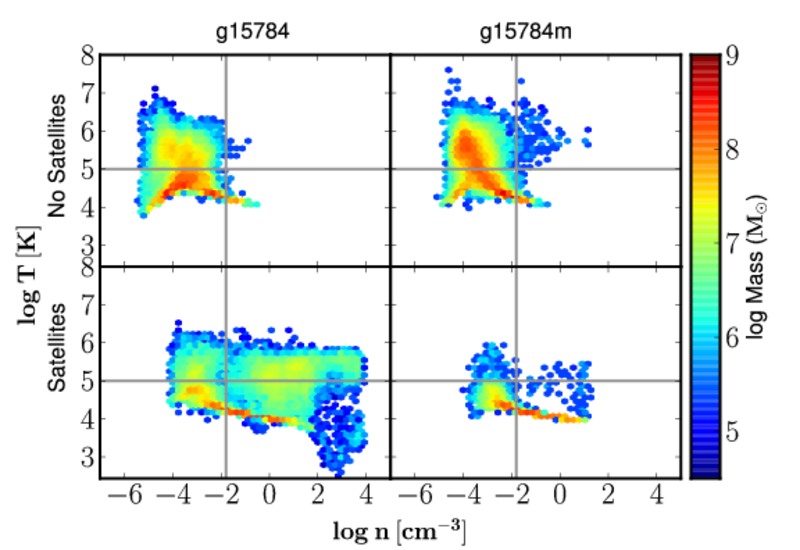}
  \caption{The same as Fig. \ref{fig:phaseshell}, but for the gas well outside (1.5-2 $r_{\mbox{vir}}$) the virial radius of the galaxy. This shows the gas phase prior to entering the galaxy. The left and right column are for g15784 and g15784m, respectively. The top row represents all gas that has not been identified inside of a satellite halo, whereas the bottom row is the gas that has been identified inside of a satellite halo. Binning is the same as Fig. \ref{fig:phaseshell}.}
  \label{fig:phaseshelligmz3}
\end{figure}

Figs \ref{fig:phaseshell} and \ref{fig:phaseshelligmz3} show phase diagrams in four radial shells of the galaxy g15784 at z = 3 for the weak and strong feedback (MAGICC) case. In Fig. \ref{fig:phaseshell}, the three rows represent shells with radial bins from 0-0.1, 0.1-0.5 and 0.5-1 r$_{\mbox{vir}}$. The left column shows the weak feedback run and the right column denotes the strong feedback run. Fig. \ref{fig:phaseshelligmz3} shows phase diagrams for the gas in the radial bin from 1.5-2 r$_{\mbox{vir}}$. The top row is for gas that has not been identified as being part of a satellite halo, and the bottom row is for gas that has been identified as part of a satellite. The columns are the same as in Fig. \ref{fig:phaseshell}.

Within the galaxy, the region around a density threshold of $0.01-0.1$ atoms cm$^{-3}$ is sparsely occupied by hot gas, indicating that gas rapidly transitions through this part of phase space.   We tracked individual particles near this density and we find that they tend to progress smoothly to higher densities without backtracking. There is thus a natural opportunity to examine accreting gas as its density increases from a low pre-virial value of less than $10^{-4}$ cm$^{-3}$ and approaches the interstellar medium (ISM) near $10^{-2}$ cm$^{-3}$.  

Similarly, gas at temperatures T $\sim 10^5$ K is relatively uncommon once densities start to rise due to the peak in cooling at this temperature.  At the intermediate densities present within the virial radius, gas tends to be hot ($> 10^5$ K), with long cooling times or cool, at $10^4$ K or below. Thus, $10^5$ K provides a natural point of division separating hot halo gas and cooler accreting gas, which takes on a bimodal distribution once within the virial radius.  This is particularly clear in the second (0.1-0.5 $r_{\mbox{vir}}$) and third (0.5-1 $r_{\mbox{vir}}$) rows of Fig.~\ref{fig:phaseshell}, showing gas that is within the virial radius but outside the star-forming ISM.

It should be noted that gas at high densities ($> 10$ cm$^{-3}$) can have effective temperatures near $10^5$ K without rapid cooling (such as in the top row of Fig.~\ref{fig:phaseshell} that includes the bulk of the ISM). For dense gas undergoing feedback, the simulation temperature represents the mean {\it effective} temperature of the ISM rather than actual gas temperatures.  Due to finite mass resolution, combined regions with hot winds, supernova (SN) ejecta and regular ISM are represented with single resolution elements.  The feedback model corrects the cooling rates to take the multiphase aspect into account but does not provide a detailed breakdown for analysis. However, for the purposes of this study, it is correct to assume that resolution elements which were previously cool and dense and then rose past $10^5$ K have undergone substantial stellar feedback. It should be noted that the effective post-feedback temperatures {and the radius to which gas is pushed} are higher for the stronger feedback (e.g. the right column of Fig.~\ref{fig:phaseshell}).

The above temperature ($10^5$ K) and density (0.01 cm$^{-3}$) values provide natural and physically motivated thresholds to classify accreting gas. Gas that accretes hot ($> 10^5$ K) has significantly longer cooling times and thus is delayed from forming into stars. For this reason, we classify all gas that accretes at a temperature above $10^5$ K as \textit{hot accretion} and gas that accretes below this temperature as \textit{cold accretion}. For gas that forms into stars, we consider two more possible categories that include the effect of feedback. If gas rises above the temperature threshold ($10^5$ K) at a density above the density threshold (0.01 cm$^{-3}$), it is considered to have received stellar/SN feedback, and is classified as \textit{hot/cold fountain} (hot/cold in this case describes its initial accretion into the galaxy). Note that the initial hot/cold classification does not distinguish \textit{how} the gas acquired its temperature. This means that the `hot' category contains both shock-heated and pre-heated gas, and the `cold' category contains both Intergalactic Medium (IGM) and ISM gas (within satellites). While this classification scheme is more simplistic than other authors, e.g. \citet{brooks09}, its simplicity and physical motivation make it relevant to observations and easy to interpret in the context of star-forming gas. The specific conditions are summarized below:

\begin{itemize}
\item \emph{cold accretion}: $T < 10^5$ K while density $n < 10^{-2}$ cm$^{-3}$;
\item \emph{hot accretion}: $T > 10^5$ K while density $n < 10^{-2}$ cm$^{-3}$;
\item \emph{fountain}: heated one or more times to $T > 10^5$ K at $n > 10^{-2}$ cm$^{-3}$:
\begin{itemize}
 \item \emph{hot fountain}:  {\it hot accretion} and {\it fountain};
 \item \emph{cold fountain}:  {\it cold accretion} and {\it fountain}.
\end{itemize}
\end{itemize}

Note that \textit{hot accretion} represents both the traditional idea of shock-heated gas, as well as gas that was heated prior to accretion by any other means. \textit{Hot fountain} represents \textit{hot accretion} gas which has been metal enriched and delayed from forming into stars via heating by stellar/SN feedback (e.g. \citealt{cox81}). {\it Cold fountain} is similar, but for {\it cold accretion} gas.

In the case of gas that first enters a smaller halo which is later incorporated into the final galaxy halo, the mode of accretion is based on its passage into the final halo. For any gas which is ejected entirely from the main halo (due to feedback or other processes) and then reaccreted at a later time, the classification is based on the first accretion.

Particle tracking was limited to the time resolution of the output files, which was $\le$~214 Myr. This is sufficient resolution to accurately determine the above categories, as \citet{brooks09} found that reducing the time between outputs from 320 to 27~Myr resulted in only an $\sim$2 per cent change in the amount of shock-heated gas. As our classification scheme is simpler, our results should not be significantly affected by time resolution of outputs.

\section{Results}
\label{sec:results}

In this section, we first present a time history of the accretion modes of all galaxies. Next, we look at how the different accretion modes feed star formation. We look at both how much mass in each mode ends up in stars, as well as how long each mode typically takes to form stars. Finally, we look at how feedback affects each accretion/formation mode.

\subsection{Gas Accretion}
\label{sec:gasaccretion}

\begin{figure*}
 \includegraphics[natwidth=900,natheight=400,width=\textwidth]{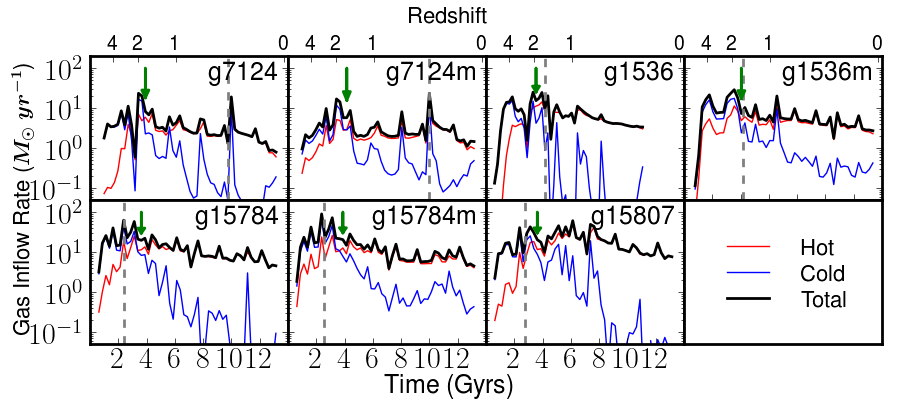}
  \caption{The gas accretion history of all galaxies.  Red represents gas that was above $10^5$ K at accretion, and blue shows the gas below $10^5$ K at accretion. Black is the total accretion rate. The vertical grey line represents the point in time at which the galaxy crosses the $10^{11.6}~M_{\odot}$ threshold. Note that the transition from hot- to cold-dominated accretion does not coincide with crossing this mass threshold, but instead occurs at a similar redshift.}
  \label{fig:gasvtime}
\end{figure*}

Gas accretion has been split specifically on the mode of accretion upon entering the halo where the gas forms stars, or in the case of no formation, upon entering the latest halo it was a part of. Whether or not it experiences feedback later is ignored at this stage.  An initial goal is to test the prediction of \citet{dekelBirnboim06} that the accretion mode transition occurs at a common mass.  We have compared against a value of $10^{11.6}~M_{\odot}$, the mass where \citet{ocvirk08} find virial radius shocking to begin.

Fig. \ref{fig:gasvtime} shows a plot of accretion rate versus time for the galaxies. On these plots, the thin red line indicates hot-mode accretion, the thin blue line indicates cold-mode accretion and the thick black line shows the total accretion rate.  All of the curves show a steady progress in time with superimposed short-term features. These features are related to substructure within the haloes. The substructures directly modulate gas inflow, particularly for colder gas. However, they also make group finding more challenging because they can cause the boundaries between groups to jump around somewhat from step to step. For this reason, we avoid placing too much significance on individual spikes in the accretion histories and focus on the longer term trends.   

The curves clearly show how important cold-mode accretion is to obtaining gas at early times. All of the galaxies show a strong cold-mode accretion rate at early times. The majority of the fuel for star formation at that time is coming from cold-mode accretion. Cold-mode accretion declines exponentially past $z \sim 2$ with strong modulation due to cold gas associated with substructure. On the other hand, hot-mode accretion rises strongly to about $z \sim 2$, and then is relatively flat with at most a gentle decline to $z=0$.

A vertical grey line has been drawn on to indicate when the galaxy crosses the predicted mass threshold of $10^{11.6}~M_{\odot}$. In all cases except for g1536-magicc, the cold-to-hot transition does not occur at the mass predicted by \citet{ocvirk08}. Rather, the transitions all seem to occur near redshift of $\sim 1.8$. In the case of g1536, we argue that even though hot accretion first takes over slightly before a redshift of 1.8, the transition is longer and ends at $z\sim 1.8$. For g1536-magicc, the transition occurs at the mass threshold predicted by \citet{ocvirk08}, but only because the galaxy reaches the mass threshold at the common redshift. Table \ref{tab:transitions} summarizes the transition redshifts and masses for each galaxy.

\begin{table}
\begin{center}
\begin{tabular}{cccc}
Galaxy & Age (Gyr)  & Redshift ($z$) & Mass ($10^{11}~M_{\odot}$)\\ \hline \hline
g7124 & 3.91 & 1.74 & 2.13 \\
g7124-magicc & 4.16 & 1.63 & 2.20 \\
g1536 & 3.54 & 1.94 & 3.05 \\
g1536-magicc & 4.05 & 1.67 & 3.87 \\
g15784 & 3.63 & 1.88 & 7.79 \\
g15784-magicc & 3.88 & 1.76 & 6.96 \\
g15807 & 3.63 & 1.88 & 7.58 \\
\hline
\end{tabular}
\end{center}
\caption[The properties of each galaxy at transition.]{The age, redshift and total mass of each galaxy when gas accretion transitions from cold to hot. The transition occurs at an average value of $z = 1.8 \pm 0.1$.}
\label{tab:transitions}
\end{table}

\subsection{Star Formation}
\label{sec:starformation}

\begin{figure*}
 \includegraphics[natwidth=900,natheight=400,width=\textwidth]{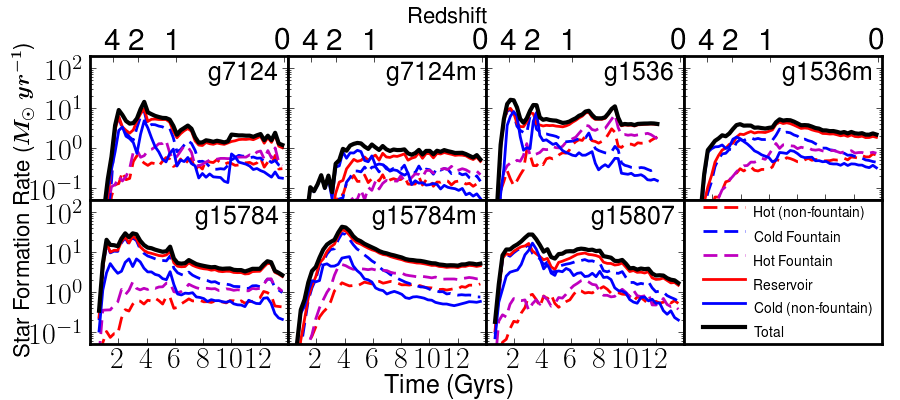}
  \caption{ Star formation history of all galaxies separated by how the parent gas accreted and evolved. The three dotted lines represent gas that jumped over the temperature threshold at some point before formation. The red dotted line shows the traditional category of gas that shock heated at accretion (without feedback). Magenta dotted represents hot gas that cooled and then experienced fountain-type feedback heating back into the reservoir before star formation. Blue dotted represents cold fountain gas: accreted in the cold mode, but heated by feedback before star formation. The solid red line shows the sum of the three dotted lines, and we have termed this {\it reservoir gas.} Finally, the solid blue line represents gas that never got above $10^5$ K before forming a star.}
  \label{fig:tempvsfr}
\end{figure*}

We next examine when stars formed versus the history of the gas they formed from. Fig. \ref{fig:tempvsfr} shows a plot of SFR separated by the mode in which the parent gas accreted. The dotted lines make up the components of what we are calling {\it reservoir gas} - hot accretion (non-fountain), hot fountain and cold fountain. This {\it reservoir} comprises all gas that was, at some point, heated above $10^5$ K, delaying that gas from forming stars. The red dotted line represents hot accretion, the blue dotted line represents cold fountain gas and the magenta dotted line shows hot fountain gas. The sum of the three is reservoir gas, shown as solid red. The solid blue line represents cold accretion that remained cold until it formed stars.

In all of the analysed galaxies, reservoir gas is the dominant fuel for star formation during most of the galaxies' histories. The important factor is when the galaxy transitions between cold and reservoir. In most cases, cold-mode accretion is very important early on, accounting for 50\%-90\% of star formation. Depending on the galaxy, reservoir gas can begin to dominate anywhere between redshifts of roughly 5 and 1. In all cases, however, late-time star formation is primarily driven by reservoir gas.

Reservoir gas itself also goes through transitions as to what accretion mode is feeding the reservoir. In all cases, cold-mode accretion dominates early on and much of this gas was also promoted to the reservoir by feedback. Thus, until redshifts of between 1 and 0.3, the cold fountain gas drives the bulk of star formation.  After this point, the hot-mode accretion begins to become the primary fuel in the reservoir.

\begin{figure*}
 \includegraphics[natwidth=800,natheight=800,width=\textwidth]{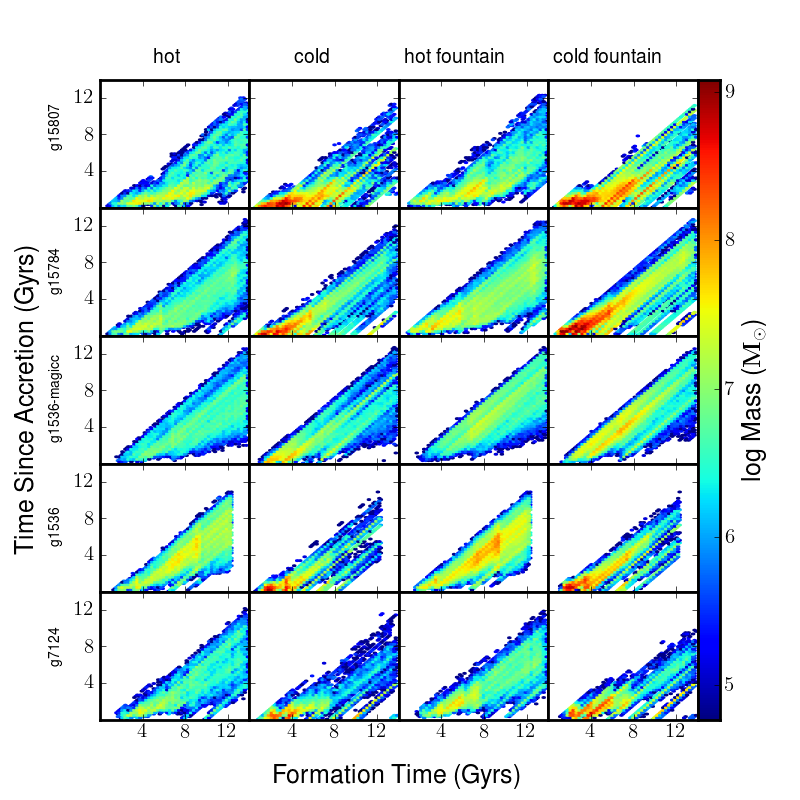}
  \caption{The joint distribution of star formation time and residence time for the parent gas in the halo. The distance above the horizontal axis indicates the delay between specific gas being accreted and it undergoing star formation.  A histogram is presented for each accretion mode presented in Fig.~\ref{fig:tempvsfr}. The difference between strong and weak feedback is similar for the cases not shown (g7124m and g15784m). Data have been binned into regular hexagons of wall-to-wall width 0.24 Gyr and height $\sqrt{3}$ width.}
  \label{fig:formationdelay}
\end{figure*}

Reservoir gas is clearly important to understand star formation, so it is useful to estimate how long gas waits in the galaxy before forming stars. This could be particularly important for galaxies that become members of groups or clusters. In that case, environmental effects might remove the reservoir without modifying the visual appearance of the galaxy.  Fig.~\ref{fig:formationdelay} is a two-dimensional histogram showing the time since the gas initially accreted on to the galaxy versus the time when the gas formed a star. Each column shows gas with different histories prior to star formation. A vertical cut at a particular time is a histogram of how long gas has been in the galaxy before forming the stars at that point in time.

In all galaxies, gas that accreted in the cold mode generally forms into stars very quickly, as one would expect. Diagonal {\it tracks} show large accretion events slowly working through all of the available gas from that event. Conversely, hot-mode gas takes up to 2 Gyr to begin forming stars once accreted, especially at late times due to the larger virial temperature and longer cooling time, and longer free-fall time.  Cold fountain gas appears very similar to cold gas at early times, but fills in the upper portions of the histogram at late times. This cold accreted gas has experienced at least one feedback event and been prevented from forming stars until later times.  Finally, hot fountain gas appears quite similar to hot gas, but fills in the very long delay section of the plot. In general, the fountain gas delays are similar to those of the non-fountain gas of the same type, but extended further due to feedback.  For example, hot feedback gas experiences substantial delays from the initial shock heating until its first entry into the ISM.  At this point, it essentially loses knowledge of its history and can undergo multiple episodes of feedback cycling prior to star formation.  This results in the longest overall delays.

Fig.~\ref{fig:formationdelay} highlights how important early accretion and large accretion events are. While much of the gas may be used immediately after the accretion event, a large percentage ends up in the reservoir and thus fuels the galaxy over the next 4-8 Gyr. In most of the tested galaxies, it is only in the past 1-2 Gyr that early cold-mode accretion has finally been used up.  Individual accretion events appear as diagonal lines in Fig.~\ref{fig:formationdelay}.  These are particularly apparent in the second panel in the top row, showing cold gas forming stars in g15807.

\subsection{The Role of Feedback}
\label{sec:roleoffeedback}

In order to get an understanding of how feedback affects the picture of cold flows and reservoirs, we included two feedback prescriptions.  The weaker feedback from the original MUGS paper resulted in large stellar fractions in the galaxies that likely conflicts with observations \citep{stinson10}.  We also included the strong feedback scheme presented in \citet{stinson13}. Neither feedback model treats individual feedback processes with detailed physical models as this is only appropriate for simulations with a fully resolved ISM.  Instead, they combine the various sources of feedback within a sub-grid framework where the feedback processes effectively behave like a locally increased pressure for the duration of the feedback. Ultimately, the energy thermalizes and cools away.  The key difference over the original feedback from MUGS is that in the stronger feedback model, feedback begins immediately and inputs considerably more energy per unit mass of new stars.  This results from higher mass loading and stronger outflows which are able to redistribute baryons both within the halo and outside it. By employing these two schemes, the current work explores the range of expected galaxy-scale impacts from stellar feedback.  In particular, we can look at how baryons enter haloes, become available for star formation and cycle back into non-star-forming reservoirs.

\begin{figure}
 \includegraphics[natwidth=800,natheight=550,width=0.5\textwidth]{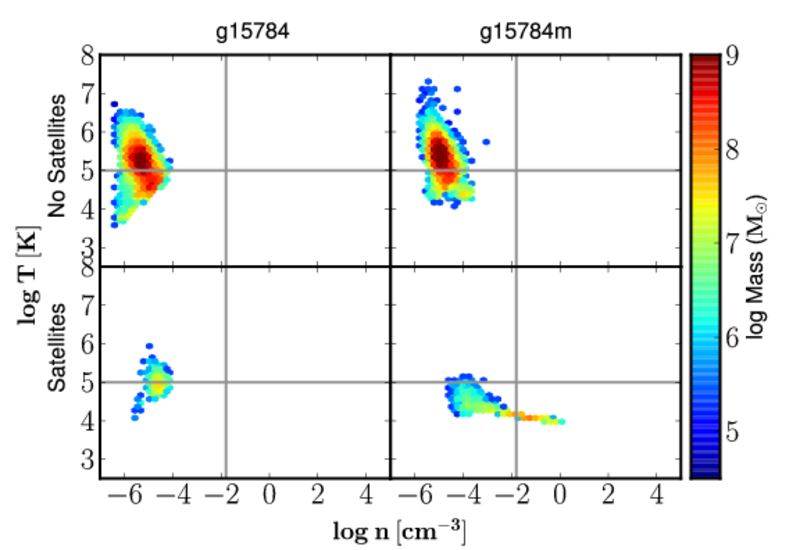}
  \caption{{The same as Fig.~\ref{fig:phaseshelligmz3}, but for $z$ = 0.5. The amount of cold gas present in satellites (bottom row) in the high-feedback case is noticeably higher, leading to a consistently higher cold accretion rate at late times in all high-feedback runs. Binning is the same as in Fig.~\ref{fig:phaseshell}.}}
  \label{fig:phaseshelligmlowz}
\end{figure}

\begin{figure*}
 \includegraphics[natwidth=900,natheight=400,width=\textwidth]{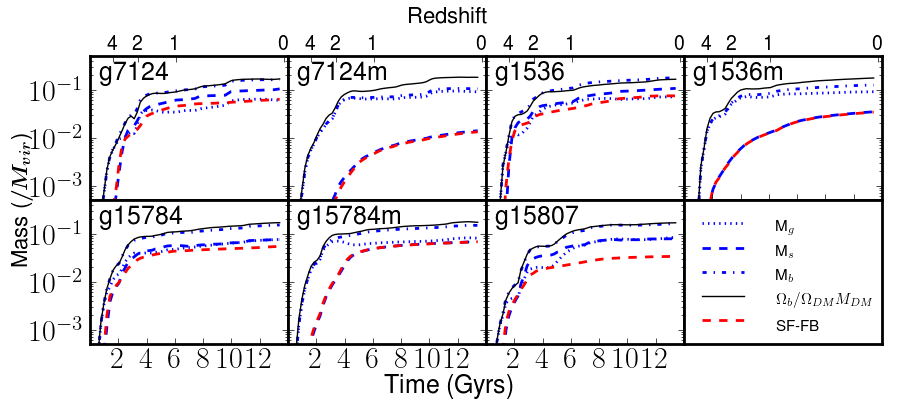}
  \caption{A plot of gas and star content over time. The dotted blue line represents gas mass, the dashed blue represents star mass and the dot-dashed blue represents baryon mass (gas + stars). The solid black shows the universal fraction of baryons, calculated as $M_{\mbox{DM}}\Omega_{\mbox{b}}$/$\Omega_{\mbox{DM}}$. The red dashed line represents the initial mass-weighted star formation minus mass lost to feedback.  The stellar content (blue dashed) may exceed the red dashed line due to stars added via mergers rather than forming within the halo.}
  \label{fig:allrates}
\end{figure*}

The effects of strong feedback can be seen in Figs.\ref{fig:gasvtime}-\ref{fig:allrates}.  Fig.~\ref{fig:gasvtime} shows that the overall gas accretion rate does not change significantly with stronger feedback. However, the relative importance of hot and cold modes does change noticeably. The different feedback history affects late accretion of gas, in particular the accretion of cooler gas at late times. This is related to the ability of gas to cool with different levels of feedback rather than delaying the accretion of specific gas. Thus the total accretion is the same but the fraction in each mode is changed by a modest amount.

Fig.~\ref{fig:phaseshelligmlowz} is the same as Fig.~\ref{fig:phaseshelligmz3}, but for a redshift of $z$ = 0.5.  It shows the phase diagrams for gas in the range of 1.5-2 r$_{\mbox{vir}}$ for particles found in satellite haloes (bottom row) and particles found outside of satellite haloes (top row). The left and right columns show the low and high feedback runs, respectively. Fig.~\ref{fig:phaseshelligmlowz} shows that there is more cold gas available for accretion in the high-feedback runs because more cold gas is present in satellite haloes.

An extreme difference is seen in the overall SFR, where at early times there is a difference of up to an order of magnitude.  Since the overall accretion rate has barely changed (Fig.~\ref{fig:gasvtime}), it is unrelated to this large change in SFR.  Since the SFR is significantly decreased, but the accretion rate is not, most of this gas must simply be unavailable to form stars, such as in the reservoir, or ejected entirely.  This is confirmed in the time delay plots in Fig.~\ref{fig:formationdelay}, where the cold fountain histogram is significantly populated over the whole history of the galaxy.  Conversely, much of the gas that formed stars with short delays (concentration of gas near the horizontal axis) has been removed with the stronger feedback (fourth row compared to third row).

% The hot and hot fountain modes stay fairly similar between the vanilla and MAGICC runs, with the main change being a simple overall decrease in the MAGICC run. However, the cold and cold fountain modes have very clearly had much of the early star formation smeared out over the length of the simulation.

Fig.~\ref{fig:allrates} shows the baryon content of the galaxies over time. The dotted and dashed blue lines represent gas and star mass within the galaxy, respectively. The dot-dashed blue line represents the baryon content of the galaxy (the sum of the blue dotted and blue dashed). The solid black line shows the universal fraction of baryons, $M_{\mbox{DM}}\Omega_{\mbox{b}}$/$\Omega_{\mbox{DM}}.$ The red dashed line represents the initial mass-weighted star formation minus mass lost in stellar feedback and SNe. In the absence of mergers, the red dashed line would equal the blue dashed line. Any discrepancy seen is due to stars obtained in mergers.

This plot illustrates the effect of feedback on the baryon content of a galaxy. In all of our weaker feedback simulations, baryon mass is essentially the same as the universal fraction (dotted line). In the case of g7124 and g1536, strong feedback reduces the amount of baryons in the galaxy by almost a factor of 2. For the larger galaxy, g15784, stronger feedback makes only a small difference in the baryon content due to the larger potential.

Additionally, the plot illustrates an interesting effect on satellites. In all weak feedback cases, there is a significant discrepancy between the red and blue dashed lines, implying that a significant number of stars in the galaxy are obtained in mergers of satellite galaxies. Conversely, the strong feedback versions do not show this discrepancy, suggesting that the mergers contain very few stars. It shows the importance of high-energy feedback in shutting down star formation in satellites. This effect is also seen indirectly in Fig.~\ref{fig:phaseshelligmlowz}, where much more cold gas is still available (has not yet been turned into stars) in accreting satellites. The lack of stellar accretion in the high-feedback runs is in agreement with authors such as \citet{behroozi13} and \citet{yang13}, who claim that the majority ($\sim$80\%) of stars in $10^{12}~M_{\odot}$ haloes form within the galaxy, and are not acquired in mergers.

\section{Discussion \& Conclusion}
\label{sec:discussion}

We have presented detailed tracking of gas histories in four galaxies with a range in masses close to the Milky Way mass from the MUGS sample \citep{stinson10} simulated with the original, fairly weak, feedback and stronger feedback described in the MAGICC papers \citep{stinson13}.  These two approaches span the range expected of stellar feedback in galaxies.

Each galaxy had all gas tracked through the history of the simulation.  Gas accretion was classified based on its temperature (hot accretion versus cold accretion). For gas that formed stars, it was further classified between hot, cold, hot fountain and cold fountain based on what happened before formation of the star.  Finally, we looked at how long gas lay within the galaxy before forming stars for each of these four classifications.

This work has revealed a detailed picture of gas accretion and star formation in galaxies.  Some aspects, such as the overall accretion at the virial radius, are insensitive to the feedback model.  At very early times, star formation is largely driven by cold gas accretion (either in filamentary flows or spherically, in the case of smaller galaxies). As the first stars begin to form, feedback starts to cycle (mostly cold-mode) gas into a hot {\it reservoir}. During this time, gas is also being added to the reservoir by classical hot accretion of gas that shocks at the virial radius.  Around a redshift of $z \sim 1.8$, cold accretion drops sufficiently for hot accretion to become more important.  Around the same time, reservoir gas begins to dominate the SFR.  Much of the cold gas added in the early stages is then being used to drive the star formation.   Different components of the reservoir are resident for different times.  Cold fountain gas, for example, is cold gas that joined the ISM and was then heated by feedback to become part of the hot reservoir.  However, it may not travel far before cooling and becoming available for star formation again.

An initial expectation for this study was that the transition from cold-mode to hot-mode accretion at the virial radius would depend strongly on halo mass, as suggested by \citet{dekelBirnboim06} and studied in \citet{ocvirk08}.  Though the precise transition mass is uncertain and may depend on other factors beyond halo mass, a basic expectation was that the transition redshift would increase with increasing halo mass, whereas we find no such trend.  As shown in Fig.~\ref{fig:gasvtime}, the transition occurs at a redshift of $z \sim 1.8$ in all cases.   Hot-mode accretion is fairly steady, whereas there is an exponential decline in cold-mode accretion, starting near $z \sim 2$. While there are no particular external factors such as sharp changes in the UV background to explain why this redshift should stand out, one could argue that less cold gas is available around this redshift due to the build up of the warm-hot intergalactic medium (WHIM; e.g. \citealt{cenOstriker99}; \citealt{dave01}; \citealt{mo05}). The idea is that the WHIM builds up during gravitational pancaking when gas shock heats while accreting on to objects with overdensities of $\sim$10, such as filaments. \citet{mo05} argue that z$\sim$2 is the point at which the cooling times of these pancakes become equal to the Hubble time, and thus gas is no longer able to cool efficiently.

It is worth noting that different studies use different definitions of cold and hot gas.  For example, \citet{keres08} label inflowing gas as cold if it has been cold for its entire history, not by its state as it enters the virial radius (as in this work).  Based on our simulation data this specific difference would not change the amount of gas in each classification very much.  However, there are some more complex characterizations (such as those using entropy changes) that would lead to substantial differences.  Other studies have also used cosmological volumes, generally with considerably lower resolution.  In this work, we concentrated on high-resolution, individual galaxies at the cost of a small sample.  We also explored more effective feedback, resulting in substantially different star formation histories compared to the majority of prior work.  These factors have the potential to significantly affect the outcomes.  In particular, the nature of the timing of the transtion from cold to hot accretion must be confirmed with high resolution studies involving a larger sample of galaxies.

%The ultraviolet background changes gradually with a broad peak around $z \sim 2.5$.   At a common redshift, the galaxies would have roughly similar densities for the accreting gas which has implications for cooling times.  We speculate that the rapid drop in density with time may play a role in allowing the galaxy to support a virial shock.  This common redshift for the transition is intriguing and should be probed in more detail in future work with a larger sample.

In this work, we draw attention to the idea of a {\it reservoir} where gas can sit within the virial radius at higher temperatures without forming stars.   Fig.~\ref{fig:formationdelay} shows that typical delays can be 2-8 Gyr, particularly for gas that is shocked to high temperatures at the virial radius.   The reservoir plays a critical role in enabling star formation at late times. The current study focuses on isolated galaxies. Denser environments will tend to first impact the reservoir and thus dramatically limit galaxies' ability to maintain star formation.  For example, as a galaxy enters a group or cluster environment, it initially loses its ability to feed the reservoir and may also have the reservoir removed by ram pressure-like effects without dramatically altering the visual appearance of the galaxy in either stars or cooler gas.

As is seen in Figs \ref{fig:tempvsfr} and \ref{fig:formationdelay}, feedback is very important for storing gas.  Large quantities of accreted gas are able to cool to ISM densities and temperatures.  Feedback can cycle this gas back into the reservoir.  This allows the gas to be used up over many Gyr.  If feedback is relatively weak, the fraction returning to the reservoir is lower and the accreted gas is used up much sooner.  If feedback is strong, galaxies may not only cycle more gas into the reservoir but also eject gas entirely (as seen in Fig.~\ref{fig:allrates}) and end up with fewer baryons and lower overall stellar content.  Another way of phrasing this is that both the mass loading and the associated velocities of the feedback gas increase with stronger feedback.  Strong feedback makes early star formation much less efficient.  In general, strong feedback is able to delay the peak of star formation.  This decouples the peak in star formation from the peak in gas accretion at $z=2-4$ and shifts it to lower redshifts, $z=1-2$.   Such a star formation history and the associated final stellar content were shown by \citet{stinson13} to be much more consistent with expectations from observations \citep[e.g.][]{moster13}.   Similar impacts on star formation from strong feedback have been seen in other codes (e.g. \citealt{scannapieco12}).   The effectiveness of feedback also depends on the mass of the galaxy.  We see that strong feedback makes a much smaller difference for our larger galaxy, g15784. In this case, the higher feedback delays the star formation peak only slightly.  However, it is still able to usefully lower the final stellar content.

\section{Acknowledgements}
\label{sec:acknowledgements}

We would like to thank the Shared Hierarchical Academic Research Computing Network (SHARCNET: www.sharcnet.ca) for use of their machines in the MUGS runs and analysis of this work. JW and HMPC thank NSERC for their support.  GS received funding from the European Research Council under the European Union's Seventh Framework Programme (FP 7) ERC Grant Agreement no. 321035.

\bibliography{WoodsEt2013MNRAS_arxiv}
\label{lastpage}

\end{document}